\documentstyle[prl,eqsecnum,aps]{revtex}

\begin{document}
\author{Jian-Qi Shen \footnote{E-mail address: jqshen@coer.zju.edu.cn}$,$ Hong-Yi Zhu \footnote{E-mail address: zhy@coer.zju.edu.cn} and Pan Chen}
\address{Center for Optical
and Electromagnetic Research, State Key Laboratory of Modern \\
Optical Instrumentation, College of Information Science and
Engineering; \\Zhejiang Institute of Modern Physics and Department
of Physics, \\Zhejiang University, Hangzhou 310027, People's
Republic of China}
\date{\today}
\title{A unified approach to exact solutions of time-dependent \\Lie-algebraic quantum systems}
\maketitle

\begin{abstract}
By using the Lewis-Riesenfeld theory and the invariant-related
unitary transformation formulation, the exact solutions of the
{\it time-dependent} Schr\"{o}dinger equations which govern the
various Lie-algebraic quantum systems in atomic physics, quantum
optics, nuclear physics and laser physics are obtained. It is
shown that the {\it explicit} solutions may also be obtained by
working in a sub-Hilbert-space corresponding to a particular
eigenvalue of the conserved generator ( {\it i. e.}, the {\it
time-independent} invariant ) for some quantum systems without
quasi-algebraic structures. The global and topological properties
of geometric phases and their adiabatic limit in time-dependent
quantum systems/models are briefly discussed.
\\ \\
PACS number(s): 03.65.-w, 03.65.Fd, 42.50.Gy
\\ \\
Shortened version \footnote {Shortened version of the title: Exact solutions of quantum systems...}

PACS \footnote {

 03.65.Fd  - Algebraic methods

 42.50.Gy  - multi-photon processes}

\end{abstract}


\section{Introduction}

Exact solutions and geometric phase
factor\cite{Berry,Vinet,Shen1,Zhu} of time-dependent spin model
have been extensively investigated by many authors\cite{Shen2,Bouchiat,Datta,Mizrahi,Gao0}.
Bouchiat and Gibbons discussed the geometric phase for the
spin-$1$
system\cite{Bouchiat}. Datta {\it et al} found the exact solution for the spin- $%
\frac{1}{2}$ system \cite{Datta} by means of the classical Lewis-Riesenfeld
theory, and Mizrahi calculated the Aharonov-Anandan phase for the spin- $\frac{1}{2}$ system
\cite{Mizrahi} in a time-dependent magnetic field. The more systematic
approach to obtaining the formally exact solutions for the spin- $j$ system
was proposed by Gao {\it et al}\cite{Gao0} who made use of the Lewis-Riesenfeld
quantum theory\cite{Lewis}. In this spin- $j$ system, the three Lie-algebraic generators
of the Hamiltonian satisfy the commutation relations of $SU(2)$ Lie algebra.
In addition to the spin model, there exist many quantum systems whose
Hamiltonian is also constructed in terms of three generators of various Lie
algebras, which we will illustrate in the following.

The invariant theory that can be applied to solutions of the
time-dependent Schr\"{o}dinger equation was first proposed by Lewis and
Riesenfeld in 1969\cite{Lewis}. This theory is appropriate for treating
the geometric phase factor. In 1991, Gao {\it et al} generalized this theory and
put forward the invariant-related unitary transformation formulation\cite{Gao1}.
Exact solutions for time-dependent systems obtained by using the generalized
invariant theory contain both the geometric phase and the dynamical phase\cite
{Gao2,Gao3,Gao4}. This formulation was developed from the Lewis-Riesenfeld's formal theory
and proven useful to the treatment of the exact solutions of
the time-dependent Schr\"{o}dinger equation and geometric phase factor. In
the present paper, based on these invariant theories we obtain exact solutions of various time-dependent
 quantum systems with the three-generator Lie-algebraic structures.

This paper is organized as follows: in Sec. 2, we set out
several quantum systems and models to illustrate the fact that
many quantum systems and models possess three-generator Lie-algebraic
structures; in Sec. 3, use is made of the invariant theories
and exact solutions of various time-dependent three-generator
systems are therefore obtained; in Sec. 4, there are some
discussions concerning the closure property of the Lie-algebraic
generators in the sub-Hilbert-space. In Sec. 5, we conclude this paper
with some remarks.

\section{The algebraic structures of various three-generator quantum systems}

In our previous work\cite{Shen} we have shown that the time-dependent
Schr\"{o}dinger equation is solvable if its Hamiltonian is constructed in
terms of the generators of a certain Lie algebra. This, therefore, implies that analyzing the algebraic
structures of Hamiltonians plays significant role in obtaining exact
solutions of the time-dependent systems. To the best of our knowledge, a large number
of quantum systems, which have three-generator Hamiltonians, have been
considered in the literature. Most of them, however, are considered only in the time-independent cases,
where the coefficients of the Hamiltonians are independent of time. In the present paper, we try to obtain
the complete set of exact solutions of all these quantum systems in the time-dependent cases.
In the following we set out these systems
and discuss the algebraic structures of their Hamiltonians.
\\ \\
(1) Spin model. The time evolution of the wavefunction of a spinning
particle in a magnetic field was studied by regarding it as a spin model\cite
{Mizrahi} whose Hamiltonian can be written

\begin{equation}
H(t)=c_{0}\{\frac{1}{2}\sin \theta \exp [-i\varphi ]J_{+}+\frac{1}{2}\sin
\theta \exp [i\varphi ]J_{-}+\cos \theta J_{3}\}  \label{eq1}
\end{equation}
with $J_{\pm }=J_{1}\pm iJ_{2}$ satisfying the commutation relations $%
[J_{3},J_{\pm }]=\pm J_{\pm },[J_{+},J_{-}]=2J_{3}.$ Analogous to
this case, in the gravitational theory of general relativity the
Hamiltonians of both the spin-gravitomagnetic
interaction\cite{Kleinert} and the spin-rotation coupling\cite
{Shen,Mashhoon1,Mashhoon2} can be constructed in terms of
$J_{+},J_{-}$, and $J_{3}$. This, therefore, means that these
interactions can be described by the spin model. It can be
verified that the investigation of the propagation of a photon
inside the noncoplanarly curved optical fiber
\cite{Chiao,Tomita,Kwiat,Ma} is also equivalent to that of a spin
model. The Hamiltonian of spin model is composed of three
generators which constitute $SU(2)$ algebra.

(2) Two-coupled harmonic oscillator. The Hamiltonian of the two-coupled
harmonic oscillator, which can describe the interaction of laser field with heat reservoirs\cite{Haken},
is of the form ( in the unit $\hbar =1$ )

\begin{equation}
H=\omega _{1}a_{1}^{\dagger }a_{1}+\omega _{2}a_{2}^{\dagger
}a_{2}+ga_{1}^{\dagger }a_{2}+g^{\ast }a_{2}^{\dagger }a_{1},  \label{eq2}
\end{equation}
where $a_{1}^{\dagger },a_{2}^{\dagger },a_{1},a_{2}$ are the
creation and annihilation operators for these two harmonic
oscillators, respectively; $g$ and $g^{\ast }$ are the coupling
coefficients and $g^{\ast }$ denotes the complex conjugation of $g$. Set $J_{+}=a_{1}^{\dagger
}a_{2},J_{-}=a_{2}^{\dagger
}a_{1},J_{3}=\frac{1}{2}(a_{1}^{\dagger }a_{1}-a_{2}^{\dagger
}a_{2}),N=\frac{1}{2}(a_{1}^{\dagger }a_{1}+a_{2}^{\dagger
}a_{2}),$ and then one may show that the generators of this
Hamiltonian represent the generators of the SU(2) subalgebra in the Weyl-Heisenberg algebra.
Since $N$ commutes with $H$, i.e., $[N,H]=0,$ we consequently say
$N$ is an invariant ( namely, it is a conserved generator whose
eigenvalue is time-independent ). In terms of $J_{\pm },J_{3}$ and
$N,$ the Hamiltonian in the expression (\ref {eq2}) can be
rewritten as follows

\begin{equation}
H=\omega _{1}(N+J_{3})+\omega _{2}(N-J_{3})+gJ_{+}+g^{\ast }J_{-}.
\label{eq3}
\end{equation}
Another interesting Hamiltonian of the two-coupled harmonic oscillator is
written in the form

\begin{equation}
H=\omega _{1}a_{1}^{\dagger }a_{1}+\omega _{2}a_{2}^{\dagger
}a_{2}+ga_{1}a_{2}+g^{\ast }a_{1}^{\dagger }a_{2}^{\dagger },  \label{eq4}
\end{equation}
which may describe the atomic dipole-dipole interaction without the rotating wave approximation\cite{Peng}. If we take $K_{+}=a_{1}^{\dagger }a_{2}^{\dagger },K_{-}=a_{1}a_{2},K_{3}=%
\frac{1}{2}(a_{1}a_{1}^{\dagger }+a_{2}^{\dagger }a_{2}),N=\frac{1}{2}%
(a_{1}^{\dagger }a_{1}-a_{2}^{\dagger }a_{2}),$ then the generators of the $%
SU(1,1)$ group are thus realized. The commutation relations are immediately
inferred as

\begin{equation}
\lbrack K_{3},K_{\pm }]=\pm K_{\pm },\quad [K_{+},K_{-}]=-2K_{3}.
\label{eq5}
\end{equation}

(3) $SU(1,1)\uplus _{s}h(4)$ Lie-algebra system. A good number of
quantum systems whose Hamiltonian is some combinations of the
generators of a Lie algebra, e.g., $SU(1,1)\uplus _{s}h(4)$
($\uplus _{s}$denotes a semidirect sum)\cite{Dattoli1,Dattoli2},
which is used to discuss both the non-Poissonian effects in a
laser-plasma scattering and the pulse propagation in a
free-electron laser\cite{Dattoli2}. The $SU(1,1)\uplus _{s}h(4)$
Hamiltonian is

\begin{equation}
H=AK_{3}+FK_{+}+F^{\ast }K_{-}+Ba^{\dagger }+B^{\ast }a+G,  \label{eq6}
\end{equation}
where $a$ and $a^{\dagger }$ are harmonic-oscillator annihilation and
creation operators, respectively.

(4) General harmonic oscillator. The Hamiltonian of the general harmonic
oscillator is given by\cite{Gao1}

\begin{equation}
H=\frac{1}{2}[Xq^{2}+Y(qp+pq)+Zp^{2}]+Fq,  \label{eq7}
\end{equation}
where the canonical coordinate $q$ and the canonical momentum $p$ satisfy
the commutation relation $[q,p]=i.$ The following three-generator Lie
algebra is easily derived

\begin{equation}
\lbrack q^{2},p^{2}]=2\{i(qp+pq)\},\quad [i(qp+pq),q^{2}]=4q^{2},\quad
[i(qp+pq),p^{2}]=-4p^{2}.  \label{eq8}
\end{equation}

(5) Charged particle moving in a magnetic field. The motion of a particle
with mass $\mu $ and charge $e$ in a homogeneous magnetic field $\vec{B}%
=(0,0,B)$ is described by the following Hamiltonian in the spherical
coordinates

\begin{equation}
H=-\frac{1}{2\mu }(\frac{\partial ^{2}}{\partial r^{2}}+\frac{2}{r}\frac{%
\partial }{\partial r}-\frac{1}{r^{2}}L^{2})+\frac{1}{8}\mu \omega ^{2}r^{2}-%
\frac{\omega }{2}L_{z}  \label{eq9}
\end{equation}
with $\omega =\frac{B}{\mu }$, where $L^{2}$ and $L_{z}$ respectively denote
the square and the third component of the angular momentum operator of the
particle moving in the magnetic field. Since both $L_{z}$ and $L^{2}$ commute with $%
H $ and thus they are called invariants, only the operators associated with $%
r$ should be taken into consideration. We can show that if the following
operators are defined

\begin{equation}
K_{1}=\mu r^{2},\quad K_{2}=-\frac{1}{\mu }(\frac{\partial ^{2}}{\partial
r^{2}}+\frac{2}{r}\frac{\partial }{\partial r}-\frac{1}{r^{2}}L^{2}),\quad
K_{3}=-2i(\frac{3}{2}+r\frac{\partial }{\partial r}),  \label{eq10}
\end{equation}
then $K_{1},K_{2}$ and $K_{3}$ form an algebra

\begin{equation}
\lbrack K_{1},K_{2}]=2iK_{3},\quad [K_{3},K_{2}]=4iK_{2},\quad
[K_{3},K_{1}]=-4iK_{1}.  \label{11}
\end{equation}
Apparently, $H$ can be rewritten in terms of the generators of this Lie
algebra.

(6) Two-level atomic coupling. The model under consideration is consisted of
two-level atom driven by the photons field\cite{Zhou}. The interaction part
of the Hamiltonian contains the transition operator $\left| 1\right\rangle
\left\langle 2\right| $ and $\left| 2\right\rangle \left\langle 1\right| $,
where $\left| 1\right\rangle $ and $\left| 2\right\rangle $ are the atomic
operators of the two-level atom. Simple calculation yields

\begin{eqnarray}
\left[ {\left| 1\right\rangle \left\langle 1\right| -\left| 2\right\rangle
\left\langle 2\right| ,\left| 1\right\rangle \left\langle 2\right| }\right]
&=&2\left| 1\right\rangle \left\langle 2\right| ,\quad
\left[ {\left|
1\right\rangle \left\langle 1\right| -\left| 2\right\rangle \left\langle
2\right| ,\left| 2\right\rangle \left\langle 1\right|}\right]
=-2\left|
2\right\rangle \left\langle 1\right| ,  \nonumber \\
\left[ {\left| 1\right\rangle \left\langle 2\right| ,\left| 2\right\rangle
\left\langle 1\right| }\right]&=&\left| 1\right\rangle \left\langle 1\right|
-\left| 2\right\rangle \left\langle 2\right| ,  \label{eq12}
\end{eqnarray}
which unfolds that the Hamiltonian contains a $SU(2)$ algebraic structure.

(7) Supersymmetric Jaynes-Cummings model. In addition to the
ordinary Jaynes-Cummings models\cite{Jaynes}, there exists a
two-level multiphoton Jaynes-Cummings model which possesses
supersymmetric structure. In this generalization of the
Jaynes-Cummings model, the atomic transitions are mediated by $k$
photons \cite{Sukumar,Sukumar2,Kien}. Singh has shown that this
model can be used to study multiple atom scattering of radiation
and multiphoton emission, absorption, and laser
processes\cite{Singh}. The Hamiltonian of this model under the
rotating wave approximation is given by

\begin{equation}
H(t)=\omega (t)a^{\dagger }a+\frac{\omega _{0}(t)}{2}\sigma
_{z}+g(t)(a^{\dagger })^{k}\sigma _{-}+g^{\ast }(t)a^{k}\sigma _{+},
\label{eq13}
\end{equation}
where $a^{\dagger }$ and $a$ are the creation and annihilation operators for
the electromagnetic field, and obey the commutation relation $\left[
a,a^{\dagger }\right] =1$; $\sigma _{\pm }$ and $\sigma _{z}$ denote the
two-level atom operators which satisfy the commutation relation $\left[
\sigma _{z},\sigma _{\pm }\right] =\pm 2\sigma _{\pm }$ . We can verify that
this model is solvable and the complete set of exact solutions can be found
by working in a sub-Hilbert-space corresponding to a particular eigenvalue
of the supersymmetric generator $N^{^{\prime }}$%
\begin{equation}
N^{^{\prime }}=\left(
\begin{array}{cc}
a^{k}(a^{\dagger })^{k} & 0 \\
0 & (a^{\dagger })^{k}a^{k}
\end{array}
\right) .  \label{eq14}
\end{equation}
It can be verified that $N^{^{\prime }}$ commutes with the Hamiltonian in (%
\ref{eq13}), and $N^{^{\prime }}$ is therefore called the
time-independent invariant. the commutation relations of its
supersymmetric Lie- algebraic structure are

\begin{eqnarray}
\left[ Q^{\dagger },Q\right] &=&\lambda _{m}\sigma _{z},\quad
\left[ N,Q\right]
=Q,\quad \left[ N,Q^{\dagger }\right] =-Q^{\dagger },  \nonumber \\
\left[ Q,\sigma _{z}\right] &=&2Q,\quad \left[ Q^{\dagger },\sigma _{z}%
\right] =-2Q^{\dagger },  \label{eq035}
\end{eqnarray}
where

\begin{equation}
N=a^{\dagger }a+\frac{k-1}{2}\sigma _{z}+\frac{1}{2},\quad
Q=(a^{\dagger })^{k}\sigma _{-},\quad Q^{\dagger }=a^{k}\sigma
_{+},  \label{eq036}
\end{equation}
and $\lambda _{m}=\frac{(m+k)!}{m!} $ denotes the eigenvalue of the time-independent invariant $%
N^{^{\prime }}$ with the eigenvalue equation being

\begin{equation}
N^{^{\prime }}{%
{\left| m\right\rangle  \choose \left| m+k\right\rangle }%
}=\lambda _{m}{%
{\left| m\right\rangle  \choose \left| m+k\right\rangle }.%
}  \label{eq0035}
\end{equation}
 By the aid of
(\ref{eq035}) and (\ref{eq036}), the Hamiltonian (\ref{eq13}) of
this supersymmetric Jaynes-Cummings model can be rewritten as

\begin{equation}
H(t)=\omega (t)N+\frac{\omega (t)-\delta (t)}{2}\sigma
_{z}+g(t)Q+g^{\ast }(t)Q^{\dagger }-\frac{\omega (t)}{2}.
\label{eq037}
\end{equation}
with $\delta (t)=k\omega (t)-\omega _{0}(t).$

Vogel and Welsch have studied the $k$-photon Jaynes-Cummings
model with coherent atomic preparation which is time-independent\cite{Vogel}%
. In the framework of the formulation presented in this paper, we can study
the totally time-dependent cases of work done by Vogel and Welsch\cite{Vogel}.

(8) Two-level atom interacting with a generalized cavity. Consider the
following Hamiltonian\cite{Yu}

\begin{equation}
H=r(A_{0})+s(A_{0})\sigma _{z}+gA_{-}\sigma _{+}+g^{\ast }A_{+}\sigma _{-}
\label{eq15}
\end{equation}
where $r(A_{0})$ and $s(A_{0})$ are well-defined real functions of $A_{0},$
and $A_{0},A_{\pm }$ satisfy the commutation relations $[A_{0},A_{\pm }]=\pm
mA_{\pm }\cite{Bonatsos}$. One can show that this Hamiltonian possesses a
three-generator algebraic structure.

(9) The interaction between a hydrogenlike atom and an external magnetic
field. This model is described by

\begin{eqnarray}
H &=&\alpha \vec{L}\cdot \vec{S}+\beta (L_{z}+2S_{z})  \nonumber \\
&=&\beta L_{z}+(\frac{1}{2}\alpha L_{z}+\beta )\sigma _{z}+\frac{1}{2}\alpha
(L_{-}\sigma _{+}+L_{+}\sigma _{-})  \label{eq16}
\end{eqnarray}
with $L_{\pm }$ $=L_{x}\pm iL_{y}.$ It is evidently seen that this form of
Hamiltonian is analogous to that in (\ref{eq15}).

(10) Coupled two-photon lasers. The Hamiltonian of coupled two-photon lasers is in fact the
combination of the two Hamiltonians of the two-coupled harmonic oscillator and general
harmonic oscillator. One can show that there exists a $SU(2)$ algebraic
structure in this model\cite{Gomes}.
\\ \\
From what has been discussed above we can draw a conclusion that a
number of typical and useful systems and models in laser physics,
atomic physics and quantum optics can be attributed to various
three-generator types. Dattoli {\it et al} have also studied the
Lie-algebraic structures and time evolutions of the most of the above
illustrative examples\cite{Dattoli}. It should be noted that most
of above systems and models in the literature were only considered in the stationary
cases where the coefficients of the Hamiltonians were totally
time-independent ( or partly time-dependent ). In the present
paper, we will further indicate in what follows that the analysis
of these algebraic structures shows the solvability of these
quantum systems. In the meanwhile we give exact solutions of the time-dependent
Schr\"{o}dinger equation of all these systems and models where all
the coefficients of the Hamiltonians are time-dependent.

\section{ Exact solutions of time-dependent Schr\"{o}dinger equation}

Time evolution of most above systems and models is governed by the
Schr\"{o}dinger equation

\begin{equation}
i\frac{\partial \left| \Psi (t)\right\rangle _{s}}{\partial t}=H(t)\left|
\Psi (t)\right\rangle _{s},  \label{eq17}
\end{equation}
where the Hamiltonian is constructed by three generators $A,B$ and $C$ and
is often given as follows

\begin{equation}
H(t)=\omega (t)\{\frac{1}{2}\sin \theta (t)\exp [-i\phi (t)]A+\frac{1}{2}%
\sin \theta (t)\exp [i\phi (t)]B+\cos \theta (t)C\}  \label{eq18}
\end{equation}
with $A,B$ and $C$ satisfying the general commutation relations of a Lie
algebra

\begin{equation}
\lbrack A,B]=nC,\quad \lbrack C,A]=mA,\quad \lbrack C,B]=-mB,  \label{eq19}
\end{equation}
where $m$ and $n$ are the structure constants of this Lie algebra. Here,
for convenience, the Hamiltonians with the three-generator Lie-algebraic
structures is parameterized to be expression (\ref{eq18}) in terms of the parameters
$\omega ,\theta $ and $\phi$. For instance , in model Hamiltonian (\ref{eq3}), the
Lie-algebraic generators, $A,B$ and $C$ in expression (\ref{eq18}), may respectively stand
for $J_{+},J_{-}$, and  $J_{z}$. Thus the  parameters $\omega ,\theta $ and
$\phi $ may be determined by

\begin{eqnarray}
g(t)&=&\frac{1}{2}\omega (t)\sin \theta (t)\exp [-i\phi (t)], \quad g^{\ast }(t)
=\frac{1}{2}\omega (t)\sin \theta (t)\exp [i\phi (t)],  \nonumber\\
\omega _{1}(t)-\omega _{2}(t)
&=&\omega (t)\cos \theta (t),
\end{eqnarray}
with $\omega (t)=\sqrt{\left[ \omega _{1}(t)-\omega _{2}(t)\right] ^{2}
+4g(t)g^{\ast }(t)}$. The same parameterizing approach is also readily applied to
the supersymmetric Jaynes-Cummings model Hamiltonian (eq037) and all other three-generator Lie-algebraic
quantum systems and models presented in this paper. Since all
simple $3$-generator algebras are either isomorphic to the algebra $sl(2,C)$
or to one of its real forms, we treat these time-dependent quantum systems
in a unified way. According to the Lewis-Riesenfeld invariant theory, an
operator $I(t)$ that agrees with the following invariant equation\cite{Lewis}

\begin{equation}
\frac{\partial I(t)}{\partial t}+\frac{1}{i}[I(t),H(t)]=0  \label{eq20}
\end{equation}
is called an invariant whose eigenvalue is time-independent, i.e.,

\begin{equation}
I(t)\left| \lambda ,t\right\rangle _{I}=\lambda \left| \lambda
,t\right\rangle _{I},\quad \frac{\partial \lambda }{\partial t}=0.
\label{eq21}
\end{equation}
It is seen from Eq. (\ref{eq20}) that $I(t)$ is the linear combination of $%
A,B$ and $C$\bigskip\ and may be generally written\qquad

\begin{equation}
I(t)=y\{\frac{1}{2}\sin a(t)\exp [-ib(t)]A+\frac{1}{2}\sin a(t)\exp
[ib(t)]B\}+\cos a(t)C,  \label{eq22}
\end{equation}
where the constant $y$ will be determined below. It should be pointed out
that it is not the only way to construct the invariants. Since the product
of two invariants also satisfies Eq. (\ref{eq20})\cite{Gao1}, there are
infinite invariants of a time-dependent quantum system. But the form in Eq. (%
\ref{eq22}) is the most convenient and useful one. Substitution of (\ref
{eq22}) into Eq.(\ref{eq20}) yields

\begin{eqnarray}
y\exp (-ib)(\dot{a}\cos a-i\dot{b}\sin a)-im\omega \lbrack \exp (-i\phi
)\cos a\sin \theta -y\exp (-ib)\sin a\cos \theta ] &=&0,  \nonumber \\
\dot{a}+\frac{ny}{2}\omega \sin \theta \sin (b-\phi ) &=&0.  \label{eq23}
\end{eqnarray}
where dot denotes the time derivative. The time-dependent parameters $a$ and
$b$ are determined by these two auxiliary equations.

It is easy to verify that the particular solution $\left| \Psi
(t)\right\rangle _{s}$ of the Schr\"{o}dinger equation can be expressed in
terms of the eigenstate $\left| \lambda ,t\right\rangle _{I}$ of the
invariant $I(t),$ namely,

\begin{equation}
\left| \Psi (t)\right\rangle _{s}=\exp [\frac{1}{i}\varphi (t)]\left|
\lambda ,t\right\rangle _{I}  \label{eq24}
\end{equation}
with
\begin{equation}
\varphi (t)=\int_{0I}^{t}\left\langle \lambda ,t^{^{\prime }}\right|
[H(t^{^{\prime }})-i\frac{\partial }{\partial t^{^{\prime }}}]\left| \lambda
,t^{^{\prime }}\right\rangle _{I}{\rm d}t^{^{\prime }}.
\end{equation}
The physical meanings of $\int_{0I}^{t}\left\langle \lambda ,t^{^{\prime
}}\right| H(t^{^{\prime }})\left| \lambda ,t^{^{\prime }}\right\rangle
_{I}{\rm d}t^{^{\prime }}$ and $\int_{0I}^{t}\left\langle \lambda ,t^{^{\prime
}}\right| -i\frac{\partial }{\partial t^{^{\prime }}}\left| \lambda
,t^{^{\prime }}\right\rangle _{I}{\rm d}t^{^{\prime }}$ are dynamical and
geometric phase, respectively.

Since the expression (\ref{eq24}) is merely a formal solution of the
Schr\"{o}dinger equation, in order to get the explicit solutions we make use
of the invariant-related unitary transformation formulation\cite{Gao1} which
enables one to obtain the complete set of exact solutions of the
time-dependent Schr\"{o}dinger equation (\ref{eq17}). In accordance with the
invariant-related unitary transformation method, the time-dependent unitary
transformation operator is often of the form

\begin{equation}
V(t)=\exp [\beta (t)A-\beta ^{\ast }(t)B]  \label{eq25}
\end{equation}
with $\beta (t)=-\frac{a(t)}{2}x\exp [-ib(t)],\quad \beta ^{\ast }(t)=-\frac{%
a(t)}{2}x\exp [ib(t)]$, where the constant, $x$, will be determined below. By making use of the Glauber formula, lengthy
calculation yields

\begin{eqnarray}
I_{V} &=&V^{\dagger }(t)I(t)V(t)=\{\frac{y}{2}\exp (-ib)\sin a\cos [(\frac{mn%
}{2})^{\frac{1}{2}}ax]-\frac{(\frac{mn}{2})^{\frac{1}{2}}}{n}\exp (-ib)\cos
a\sin [(\frac{mn}{2})^{\frac{1}{2}}ax]\}A  \nonumber \\
&&+\{\frac{y}{2}\exp (ib)\sin a\cos [(\frac{mn}{2})^{\frac{1}{2}}ax]-\frac{(%
\frac{mn}{2})^{\frac{1}{2}}}{n}\exp (ib)\cos a\sin [(\frac{mn}{2})^{\frac{1}{%
2}}ax]\}B  \nonumber \\
&&+\{\cos a\cos [(\frac{mn}{2})^{\frac{1}{2}}ax]+\frac{(\frac{mn}{2})^{\frac{%
1}{2}}}{m}y\sin a\sin [(\frac{mn}{2})^{\frac{1}{2}}ax]\}C.
\end{eqnarray}
It can be easily seen that when $y$ and $x$ are taken to be
\begin{equation}
y=\frac{m}{(\frac{mn}{2})^{\frac{1}{2}}},\quad x=\frac{1}{(\frac{mn}{2})^{%
\frac{1}{2}}},
\end{equation}
one may derive that $I_{V}=C$, which is time-independent. Thus the
eigenvalue equation of the time-independent invariant $I_{V}$ may be written in
the form

\begin{equation}
I_{V}\left| \lambda \right\rangle =\lambda \left| \lambda \right\rangle
,\quad \left| \lambda \right\rangle =V^{\dagger }(t)\left| \lambda
,t\right\rangle _{I}.  \label{eq27}
\end{equation}
Under the transformation $V(t),$ the Hamiltonian $H(t)$ can be changed into

\begin{eqnarray}
H_{V}(t) &=&V^{\dagger }(t)H(t)V(t)-V^{\dagger }(t)i\frac{\partial V(t)}{%
\partial t}  \nonumber \\
&=&\{\omega \lbrack \cos a\cos \theta +\frac{(\frac{mn}{2})^{\frac{1}{2}}}{m}%
\sin a\sin \theta \cos (b-\phi )]+\frac{\dot{b}}{m}(1-\cos a)\}C
\label{eq28}
\end{eqnarray}
by the aid of Baker-Campbell-Hausdorff formula\cite{Wei}

\begin{equation}
V^{\dagger }(t)\frac{\partial }{\partial t}V(t)=\frac{\partial }{\partial t}%
L+\frac{1}{2!}[\frac{\partial }{\partial t}L,L]+\frac{1}{3!}[[\frac{\partial
}{\partial t}L,L],L]+\frac{1}{4!}[[[\frac{\partial }{\partial t}%
L,L],L],L]+\cdots
\end{equation}
with $V(t)=\exp [L(t)].$ Hence, with the help of Eq.(\ref{eq24}) and Eq.(\ref
{eq27}), the particular solution of the Schr\"{o}dinger equation is obtained

\begin{equation}
\left| \Psi (t)\right\rangle _{s}=\exp [\frac{1}{i}\varphi (t)]V(t)\left|
\lambda \right\rangle  \label{eq29}
\end{equation}
with the phase

\begin{eqnarray}
\varphi (t) &=&\int_{0}^{t}\left\langle \lambda \right| [V^{\dagger
}(t^{^{\prime }})H(t^{^{\prime }})V(t^{^{\prime }})-V^{\dagger }(t^{^{\prime
}})i\frac{\partial }{\partial t^{^{\prime }}}V(t^{^{\prime }})]\left|
\lambda \right\rangle {\rm d}t^{^{\prime }}=\varphi _{d}(t)+\varphi _{g}(t)
\nonumber \\
&=&\lambda \int_{0}^{t}\{\omega \lbrack \cos a\cos \theta +\frac{(\frac{mn}{2%
})^{\frac{1}{2}}}{m}\sin a\sin \theta \cos (b-\phi )]+\frac{\dot{b}}{m}%
(1-\cos a)\}{\rm d}t^{^{\prime }},  \label{eq30}
\end{eqnarray}
where the dynamical phase is $\varphi _{d}(t)=\lambda \int_{0}^{t}\omega
\lbrack \cos a\cos \theta +\frac{(\frac{mn}{2})^{\frac{1}{2}}}{m}\sin a\sin
\theta \cos (b-\phi )]{\rm d}t^{^{\prime }}$ and the geometric phase is $\varphi
_{g}(t)=\lambda \int_{0}^{t}\frac{\dot{b}}{m}(1-\cos a){\rm d}t^{^{\prime }}.$ It
is seen that the former phase is related to the coefficients of the
Hamiltonian such as $\omega ,\cos \theta ,\sin \theta ,$ etc., whereas the
latter is not immediately related to these coefficients. If the parameter $a$
is taken to be time-independent, $\varphi _{g}(T)=\lambda \int_{0}^{T}\frac{%
\dot{b}}{m}(1-\cos a){\rm d}t^{^{\prime }}=\frac{\lambda }{m}[2\pi (1-\cos a)]$
where $2\pi (1-\cos a)$ is an expression for the solid angle over the
parameter space of the invariant. It is of interest that $\frac{\lambda }{m}%
[2\pi (1-\cos a)]$ is equal to the magnetic flux produced by a magnetic monopole
( and the gravitomagnetic monopole ) of
strength $\frac{\lambda }{4\pi m}$ existing at the origin of the parameter
space\cite{Gravshen}. This, therefore, implies that geometric phase differs from dynamical
phase and it involves the global and topological properties of the time
evolution of a quantum system. This fact indicates the geometric and
topological meaning of $\varphi _{g}(t).$

Here we briefly concern ourselves with the model Hamiltonians such
as (\ref {eq3}) and (\ref {eq4}) with the conserved generator (
{\it i. e.}, the time-independent invariant of which the
eigenvalue is time-independent ). For the model Hamiltonian (\ref
{eq3}), it can be rewritten as follows
\begin{equation}
H(t)=g(t)J_{+}+g^{\ast }(t)J_{-}+\left[ \omega _{1}(t)-\omega _{2}(t)\right] J_{3}+\left[ \omega _{1}(t)
+\omega _{2}(t)\right] N
\end{equation}
with $N$ $=\frac{1}{2}\left( a_{1}^{\dagger }a_{1}+a_{2}^{\dagger
}a_{2}\right) $ being the time-independent invariant that
satisfies the commutation relation $\left[ N,H(t)\right] =0$ (
{\it i. e.}, $N$ commutes with the time-dependent Hamiltonian
$H(t)$ )$.$ Since $N$ \ is an invariant, the eigenvalue of $N$ may
be $\frac{1}{2}\left( n_{1}+n_{2}\right) $, where $n_{1}$ and
$n_{2}$ denote the eigenvalue of $a_{1}^{\dagger }a_{1}$ and
$a_{2}^{\dagger }a_{2}$, respectively. In Sec. 4, we will show
that a generalized quasialgebra can be found by working in a
sub-Hilbert-space corresponding to a particular eigenvalue,
$\frac{1}{2}\left( n_{1}+n_{2}\right) $, of the time-independent
invariant $N$, where $N$ can be replaced with the particular
eigenvalue, $\frac{1}{2}\left( n_{1}+n_{2}\right) $, in the
Lie-algebraic commutation relations. We thus rewrite the model
Hamiltonian (\ref {eq3}) as follows
\begin{equation}
H(t)=g(t)J_{+}+g^{\ast }(t)J_{-}+\left[ \omega _{1}(t)-\omega _{2}(t)\right] J_{3}+\frac{1}{2}\left( n_{1}+n_{2}\right)
 \left[ \omega _{1}(t)+\omega _{2}(t)\right] .
\end{equation}
It is easily seen that this form of the Hamiltonian is different from that in (\ref {eq18}) only by a time-dependent
$c$- numbers $\frac{1}{2}\left( n_{1}+n_{2}\right) \left[ \omega _{1}(t)+\omega _{2}(t)\right] ,$ which
contributes only a time-dependent dynamic phase factor, $\exp \left\{ \int_{0}^{t}\frac{1}{2}\left( n_{1}
+n_{2}\right) \left[ \omega _{1}(t^{\prime })+\omega _{2}(t^{\prime })\right] {\rm d}t^{\prime }\right\} ,$
to the particular solution of the time-dependent Schr\"{o}dinger equation. So,
the method presented above can be readily applied to the model Hamiltonian (\ref {eq3}).
In the same fashion, the model Hamiltonian (\ref {eq4})
can be rewritten

\begin{equation}
H(t)=g(t)K_{-}+g^{\ast }(t)K_{+}+\left[ \omega _{1}(t)+\omega _{2}(t)\right] K_{3}+\left[ \omega _{1}(t)-\omega _{2}(t)\right]
N-\frac{1}{2}\left[ \omega _{1}(t)+\omega _{2}(t)\right] \label{eq300}
\end{equation}
with $N$ $=\frac{1}{2}\left( a_{1}^{\dagger }a_{1}-a_{2}^{\dagger }a_{2}\right) $
being the time-independent invariant that commutes with this time-dependent Hamiltonian $H(t)$ (\ref {eq300}).
Since $N$ is an invariant, in the sub-Hilbert-space corresponding to the particular eigenvalue,
$\frac{1}{2}\left( n_{1}-n_{2}\right) $, of $N$, the Hamiltonian (\ref {eq4}) may be rewritten as follows

\begin{equation}
H(t)=g(t)K_{-}+g^{\ast }(t)K_{+}+\left[ \omega _{1}(t)+\omega _{2}(t)\right] K_{3}+\frac{1}{2}
\left( n_{1}-n_{2}\right) \left[ \omega _{1}(t)-\omega _{2}(t)\right]
-\frac{1}{2}\left[ \omega _{1}(t)+\omega _{2}(t)\right] ,
\end{equation}
which differs from the Hamiltonian (\ref {eq18}) only by a
time-dependent $c$- numbers $\frac{1}{2}\left( n_{1}-n_{2}\right)
\left[ \omega _{1}(t)-\omega _{2}(t)\right] -\frac{1}{2}\left[
\omega _{1}(t)+\omega _{2}(t)\right] $ that also contributes only
a time-dependent dynamic phase factor to the particular solution
of the time-dependent Schr\"{o}dinger equation. Hence we can
obtain the solutions of this class of model with the conserved
generator ( {\it i. e.}, the time-independent invariant ) by
working in a sub-Hilbert-space corresponding to a particular
eigenvalue of the time-independent invariant. Maamache informed
the authors of that he had obtained the exact solutions of these
two quantum models\cite{Maamache} also by using the invariant
formulation. But there is no so-called conserved generator in
Maamache's model Hamiltonians. It is verified that when the
general results Eq. (\ref{eq29}) and (\ref{eq30}) ( deducting the
dynamic phase factor associated with the conserved generators )
are applied to these two model, the solutions obtained are in
complete agreement with Maamache's results. The more complicated
cases such as the supersymmetric Jaynes-Cummings model whose
Hamiltonian possesses the conserved generator are discussed in
more detail in the next section.
\\ \\
To conclude this section, we briefly discuss the concepts of the
exact solution and the explicit solution. The expression
(\ref{eq29}) is a particular exact solution corresponding to the
eigenvalue $\lambda $ of the invariant, and the general solutions
of the time-dependent Schr\"{o}dinger equation are therefore
easily obtained by using the linear combinations of all these
particular solutions. Generally speaking, in Quantum Mechanics,
solution with chronological-product operator ( time-order operator
) $P$ is often called the formal solution. In the present paper,
however, the solution of the Schr\"{o}dinger equation governing a
time-dependent system is sometimes called the explicit solution,
for reasons that the solution does not involve time-order
operator. But, on the other hand, by using the Lewis-Riesenfeld
invariant theory, there always exist some time-dependent
parameters, {\it e. g.}, $a(t)$ and $b(t)$ in this paper which are
determined by the auxiliary equations (\ref{eq23}). According to
the traditional practice, when employed in experimental analysis
and compared with experimental results, these nonlinear auxiliary
equations should be solved often by means of numerical
calculation. From above viewpoints, the concept of explicit
solution is understood in a relative sense, namely, it can be
considered the explicit solution when compared with the time-evolution operator $%
U(t)=P\exp [\frac{1}{i}\int_{0}^{t}H(t^{^{\prime }}){\rm d}t^{^{\prime }}]$
involving the time-order operator, $P$; whereas, it cannot be considered
completely the explicit solution for it is expressed in terms of some
time-dependent parameters, which should be obtained via the auxiliary
equations. Hence, conservatively speaking, the solution of the
time-dependent system presented in the paper is often regarded as the exact solution rather than
the explicit solution.

\section{Exact solutions obtained in the sub-Hilbert-space\qquad}

In the previous section, we obtain exact solutions of some time-dependent
three-generator systems and models possessing the three-generator Lie-algebraic
structures by using these invariant theories. In what
follows there exists the problem of the closure property of the Lie-algebraic generators in the
sub-Hilbert-space, which should be further discussed.

The generalized invariant theory can only be applied to the treatment of the
system for which there exists the quasialgebra defined in Ref.\cite{Mizrahi}%
. Unfortunately, it is seen from (\ref{eq15}) and (\ref{eq16})
that there is no such quasialgebra in the Hamiltonians of example
(7)( {\it i. e.}, the supersymmetric Jaynes-Cummings model ) and
example (8)( {\it i. e.}, the two-level atom interacting with a
generalized cavity ). In order to solve these two models, we
generalize the method that has been used for finding the dynamical
algebra $O(4)$ of the hydrogen atom to treat this type of
time-dependent models. In the case of hydrogen, the dynamical
algebra $O(4)$ was found by working in the sub-Hilbert-space
corresponding to a particular eigenvalue of the Hamiltonian
\cite{Schiff}. In this paper, we will show that a generalized
quasialgebra can also be found by working in a sub-Hilbert-space
corresponding to a particular eigenvalue of the operator $\Delta
=A_{0}+m\frac{1+\sigma _{z}}{2}$ in the time-dependent model of
two-level atom interacting with a generalized cavity. This
generalized quasialgebra enables one to obtain the complete set of
exact solutions for the Schr\"{o}dinger equation. It is readily
verified that the operator $\Delta $ commutes with $H(t)$ and is
therefore a time-independent invariant according to Eq.
(\ref{eq20}). In order to unfold the algebraic structure of the
Hamiltonian (\ref {eq15}), the following three operators are
defined\cite{Yu}

\begin{equation}
\Sigma _{1}=\frac{1}{2[\chi (\Delta )]^{\frac{1}{2}}}(A_{-}\sigma
_{+}+A_{+}\sigma _{-}),\quad \Sigma _{2}=\frac{i}{2[\chi (\Delta )]^{\frac{1%
}{2}}}(A_{+}\sigma _{-}-A_{-}\sigma _{+}),\quad \Sigma _{3}=\frac{1}{2}%
\sigma _{z},  \label{eq034}
\end{equation}
where $\chi =\left\langle n\right| A_{+}A_{-}\left| n\right\rangle
,\left| n\right\rangle $ denotes the eigenstates of $A_{0}.$ It is
easy to see all these operators commute with $\Delta $ and the
quasialgebra $\{H,\Sigma _{1},\Sigma _{2},\Sigma _{3}\}$ is thus
found. This type of time-dependent models is therefore proved
solvable by working in a sub-Hilbert-space corresponding to the
eigenstates of the time-independent invariant. \qquad
\\ \\
As an illustrative example, we consider the supersymmetric
 multiphoton Jaynes-Cummings model by means of the
invariant theories in the sub-Hilbert-space. In accordance with
the Lewis-Riesenfeld invariant theory, the invariant $I(t)$ is often of the form

\begin{equation}
I(t)=c(t)Q^{\dagger }+c^{\ast }(t)Q+b(t)\sigma _{z}
\label{eq0037}
\end{equation}
where $c^{\ast }(t)$ is the complex conjugation of $c(t),$ and
$b(t)$ is
real. Substitution of the expressions (\ref{eq0037}) and (\ref{eq037}) for $%
I(t) $ and $H(t)$ into Eq. (\ref{eq21}) leads to the following set
of auxiliary equations

\begin{eqnarray}
\dot{c}-\frac{1}{i}[c\delta +2bg] &=&0,\quad \dot{c}^{\ast }+\frac{1}{i}%
[c^{\ast }\delta +2bg^{\ast }]=0,  \nonumber \\
\stackrel{.}{b}+\frac{1}{i}\lambda _{m}(c^{\ast }g-cg^{\ast })
&=&0, \label{eq038}
\end{eqnarray}
where dot denotes the time derivative. The three time-parameters
$c,c^{\ast } $ and $b$ in $I(t)$ are determined by these three
auxiliary equations.

This time-dependent model can be exactly solved by using the
invariant-related unitary transformation formulation where the
unitary transformation operator is of the form

\begin{equation}
V(t)=\exp [\beta (t)Q-\beta ^{\ast }(t)Q^{\dagger }].
\label{eq0381}
\end{equation}
with $\beta ^{\ast }(t)$ being the complex conjugation of $\beta
(t).$ With the help of the commutation relations (\ref{eq035}), it
can be found that, by the complicated and lengthy computations, if
$\beta (t)$ and $\beta ^{\ast }(t)$ satisfy the following
equations

\begin{equation}
\sin (4\beta \beta ^{\ast }\lambda
_{m})^{\frac{1}{2}}=\frac{\lambda
_{m}(c\beta ^{\ast }+c^{\ast }\beta )}{(4\beta \beta ^{\ast }\lambda _{m})^{%
\frac{1}{2}}},\quad \cos (4\beta \beta ^{\ast }\lambda
_{m})^{\frac{1}{2}}=b, \label{eq0310}
\end{equation}
a time-independent invariant can be obtained as follows

\begin{equation}
I_{V}\equiv V^{\dagger }(t)I(t)V(t)=\sigma _{z}.  \label{eq0311}
\end{equation}
From Eq. (\ref{eq0310}), we substitute the time-dependent
parameters $\theta $ and $\phi $ for $c,c^{\ast }$ and $b$ in
$I(t)$ for simplicity and convenience, and the results are

\begin{eqnarray}
\beta &=&-\frac{\frac{\theta }{2}\exp (-i\phi )}{\lambda _{m}^{\frac{1}{2}}}%
,\quad \beta ^{\ast }=-\frac{\frac{\theta }{2}\exp (i\phi )}{\lambda _{m}^{%
\frac{1}{2}}},  \nonumber \\
c &=&-\frac{\sin \theta \exp (-i\phi )}{\lambda
_{m}^{\frac{1}{2}}},\quad c^{\ast }=-\frac{\sin \theta \exp (i\phi
)}{\lambda _{m}^{\frac{1}{2}}}. \label{eq0312}
\end{eqnarray}
Thus, the invariant $I(t)$ in (\ref{eq0037}) can be rewritten

\begin{equation}
I(t)=-\frac{\sin \theta }{\lambda _{m}^{\frac{1}{2}}}[\exp (-i\phi
)Q+\exp (i\phi )Q^{\dagger }]+\cos \theta \sigma _{z}.
\label{eq0313}
\end{equation}
In the meanwhile, under the unitary transformation (\ref{eq0381}),
the Hamiltonian (\ref{eq037}) can be transformed into

\begin{eqnarray}
H_{V}(t) &\equiv &V^{\dagger }(t)H(t)V(t)-V^{\dagger }(t)i\frac{\partial }{%
\partial t}V(t)  \nonumber \\
&=&\omega N-\frac{\omega }{2}+\{\frac{\omega }{2}(1-\cos \theta )-\frac{1}{2}%
\lambda _{m}^{\frac{1}{2}}[g\exp (i\phi )+g^{\ast }\exp (-i\phi
)]\sin
\theta +  \nonumber \\
&&+\frac{\omega -\delta }{2}\cos \theta -\frac{\dot{\phi
}}{2}(1-\cos \theta )\}\sigma _{z}  \label{eq0314}
\end{eqnarray}
The eigenstates of $\sigma _{z}$ corresponding to
the eigenvalue $\sigma =+1$ and $\sigma =-1$ are ${%
{1  \choose 0}%
}$ and ${%
{0  \choose 1}%
},$ and the eigenstate of $N^{^{\prime }}$ is $%
{\left| m\right\rangle  \choose \left| m+k\right\rangle}%
$ in terms of (\ref{eq0035}). From Eq. (\ref{eq28}), (\ref{eq29}),
(\ref {eq30}), we obtain two particular solutions of the
time-dependent Schr\"{o}dinger equation of the time-dependent
TLMJC model, which are written in the forms

\begin{equation}
\left| \Psi _{m,\sigma =+1}(t)\right\rangle =\exp \{\frac{1}{i}\int_{0}^{t}[%
\dot{\varphi}_{d,\sigma =+1}(t^{^{\prime
}})+\dot{\varphi}_{g,\sigma
=+1}(t^{^{\prime }})]{\rm d}t^{^{\prime }}\}V(t){%
{\left| m\right\rangle  \choose 0}%
}  \label{eq0315}
\end{equation}
with
\begin{eqnarray}
\dot{\varphi}_{d,\sigma =+1}(t^{^{\prime }})
&=&(m+\frac{k}{2})\omega (t^{^{\prime }})-\frac{1}{2}\lambda
_{m}^{\frac{1}{2}}\{g(t^{^{\prime }})\exp [i\phi (t^{^{\prime
}})]+g^{\ast }(t^{^{\prime }})\exp [-i\phi
(t^{^{\prime }})]\}\sin \theta (t^{^{\prime }})  \nonumber \\
&&-\frac{\delta (t^{^{\prime }})}{2}\cos \theta (t^{^{\prime }})
\end{eqnarray}
and
\begin{equation}
\dot{\varphi}_{g,\sigma =+1}(t^{^{\prime
}})=-\frac{\dot{\phi}(t^{^{\prime }})}{2}[1-\cos \theta
(t^{^{\prime }})];  \label{eq0317}
\end{equation}
and

\begin{equation}
\left| \Psi _{m,\sigma =-1}(t)\right\rangle =\exp \{\frac{1}{i}\int_{0}^{t}[%
\dot{\varphi}_{d,\sigma =-1}(t^{^{\prime
}})+\dot{\varphi}_{g,\sigma
=-1}(t^{^{\prime }})]{\rm d}t^{^{\prime }}\}V(t){%
{0 \choose \left| m+k\right\rangle }%
}  \label{eq0318}
\end{equation}
with
\begin{eqnarray}
\dot{\varphi}_{d,\sigma =-1}(t^{^{\prime }})
&=&(m+\frac{k}{2})\omega (t^{^{\prime }})+\frac{1}{2}\lambda
_{m}^{\frac{1}{2}}\{g(t^{^{\prime }})\exp [i\phi (t^{^{\prime
}})]+g^{\ast }(t^{^{\prime }})\exp [-i\phi
(t^{^{\prime }})]\}\sin \theta (t^{^{\prime }})  \nonumber \\
&&+\frac{\delta (t^{^{\prime }})}{2}\cos \theta (t^{^{\prime }})
\label{eq0319}
\end{eqnarray}
and

\begin{equation}
\dot{\varphi}_{g,\sigma =-1}(t^{^{\prime }})=\frac{\dot{\phi}(t^{^{\prime }})%
}{2}[1-\cos \theta (t^{^{\prime }})].  \label{eq0320}
\end{equation}
It should be noted that the above approach to the time-dependent
Jaynes-Cummings model is also appropriate for treating the periodic decay
and revival of some multiphoton-transitions models, which has been
investigated by Sukumar and Buck\cite{Sukumar2}.

It is readily verified that in the sub-Hilbert-space corresponding to a
particular eigenvalue of the conserved generator (the time-independent
invariant), the Hamiltonian of original two-level Jaynes-Cummings model
( mono-photon case )\cite{Jaynes} possesses the $SU(2)$ Lie-algebraic
structure, and three-level two-mode mono-photon model possesses the $SU(3)$
structure. The solution of the time-dependent case of $SU(2)$
Jaynes-Cummings model is easily obtained by taking the number of photons
mediating in the process of atomic transitions $k=1.$ Since Shumovsky {\it et al}
have considered the three-level two-mode multiphoton Jaynes-Cummings model
\cite{Shumovsky} whose Hamiltonian is time-independent, it is also of
interest to exactly solve the time-dependent supersymmetric three-level
two-mode multiphoton Jaynes-Cummings model by means of these invariant theories.

\section{Concluding remarks}
In the present paper:
\\ \\
(1) On the basis of the fact that all simple three-generator algebras
are either isomorphic to the algebra $sl(2,C)$ or to one of its
real forms, exact solutions of the time-dependent Schr\"{o}dinger
equation of all three-generator systems and models in quantum
optics, nuclear physics, solid state physics, molecular and atomic
physics as well as laser physics are provided by making use of both the
Lewis-Riesenfeld invariant theory and the invariant-related
unitary transformation formulation. We uses the unitary transformation and obtain the
explicit expression for the time-evolution operator, instead of
the formal solution that is related to the chronological product.

(2) Since it appears only in systems with time-dependent
Hamiltonian, the geometric phase factor would be easily studied if
the exact solutions of time-dependent systems had been obtained.
In the adiabatic limit, {\it i. e.}, if the parameter $a$ is taken
to be time-independent, then the geometric phase in a cycle
associated with $b(t)$ can be rewritten as $\varphi
_{g}(T)=\frac{\lambda }{m}[2\pi (1-\cos a)]$, where $2\pi (1-\cos
a)$ is an expression for the solid angle over the parameter space
of the invariant. It is well known that this phase is just the
Berry's phase ({\it i. e.}, Berry's
 non-integral phase), which is found by Berry in the quantum
adiabatic process in 1984\cite{Berry}. But in this paper, we obtain the non-adiabatic non-cyclic geometric phase
in time-dependent quantum systems, namely, Berry's phase is only the particular case of ours presented in this paper.
In view of above discussions, the invariant-related unitary transformation
formulation is a useful tool for treating the geometric phase
factor and the time-dependent Schr\"{o}dinger equation. This
formulation replaces the eigenstates of the time-dependent
invariants with those of the time-independent invariants through
the unitary transformation and thus obtain the explicit solutions, rather than the formal solutions
associated with the chronological product, of time-dependent quantum systems.

(3) It is known to us that the time-dependent Schr\"{o}dinger
equation can be solved if its Hamiltonian is constructed in terms
of the generators of a certain Lie algebra. For some quantum
systems whose time-dependent Hamiltonians possess no
quasialgebraic structures, we show that the exact solutions can
also be obtained by working in a sub-Hilbert-space corresponding
to a particular eigenvalue of the conserved generator ( {\it i.
e.}, the time-independent invariant that commutes with the
time-dependent Hamiltonian ). In Sec. 4, we obtain the complete
set of exact solutions of the time-dependent supersymmetric
multiphoton Jaynes-Cummings model in the sub-Hilbert-space
corresponding to the time-independent invariant $N^{^{\prime }}$.
Apparently, the method presented in this paper is also applicable
to the algebraic structure whose number of generators is more than
three. Additionally, it should be pointed out that the
time-dependent Schr\"{o}dinger equation is often considered in the
literature, whereas less attention is paid to the time-dependent
Klein-Gordon equation. Since it can govern the time evolution of
some scalar fields, we think that it gets less attentions than it
deserves. Work in this direction is under consideration and will
be published elsewhere.
\\ \\
{\bf Acknowledgements}  This project is supported in part by the
National Natural Science Foundation of China under the project No.
$90101024$ and $30000034$. The authors thank Xiao-Chun Gao for
helpful proposals concerning the Lie-algebraic generators in
sub-Hilbert-space.

\end{document}